\documentclass{icontop}

\begin{document}

\title{Preliminary Design of ARIES-Devasthal Faint Object Spectrograph and Camera}

\author{Soumen Mondal, Ramakant S. Yadav \and Mahendra Singh }

\institute{Aryabhatta Research Institute of observational Sciences (ARIES),\\ Manora Peak, Nainital -263 129, India \\ email: soumen@aries.ernet.in }
 
\maketitle

 \begin{abstract}

We present here the preliminary  design of ARIES-Devasthal Faint Object Spectrograph and Camera (ADFOSC), which is a multi-mode instrument for both imaging  and spectroscopy. ADFOSC is the first-generation instrument to be mounted at the axial port of the Cassegrain focus on our new  3.6m optical telescope to be installed at Devasthal, Nainital. The main design goals of the instrument are : the instrument will have capability of broad- and narrow-band imaging, low-medium resolution spectroscopy, and imaging  polarimetry. The operating wavelength range will be from 360 to 1000 $nm$ and the instrument will have remote-control capability.
 
\end{abstract}

\section{Introduction}

Aryabhatta Research Institute of Observational Sciences (ARIES) will install a 3.6$m$ optical telescope at the site of Devasthal near Nainital in India. Devasthal site is one of the best  Astronomical observing site in India \cite{sagar}. To achieve  the scientific goals of the telescope,  the Faint Object Spectrograph and Camera (FOSC) type instrument will be one of the first generation instruments to the axial port of the Cassegrain focus on the 3.6m Devasthal Optical telescope (DOT). Considering its' operating flexibility and sensitivity, many FOSC-type instruments were built for the medium to large class telescopes in the World, e.g. EFOSC2 for ESO 3.6m telescope \cite{eso}, FOCAS for 8.2m Subaru telescope \cite{focas1,focas2}, GMOS for 8.1m Gemini telescope \cite{gmos1,gmos2} and FORS for 8m ESO VLT \cite{fors1}.

ARIES-Devasthal Faint Object Spectrograph and Camera (ADFOSC) is a multi-mode instrument for both imaging, imaging   polarimetry/spectro-polarimetry and spectroscopy. The FOSC-type multi-mode instrument allows us to carry out various scientific programs in an efficient way due to their observing flexibility. Some of the scientific programs like studies of faint objects, nebulae and star clusters in the Galaxy, individual objects in nearby galaxies, the internal structure of the galaxy, spectroscopic studies of active galactic nuclei, high redshift quasars and Giant Radio sources could be possible with this kind of the backend instrument on the new 3.6m Devasthal Optical Telescope (DOT) at Devasthal, Nainital.  

There are a number of design requirements and goals of the instrument. These are: (i) the instrument should be capable of both direct good quality imaging, polarimetric  and low-medium resolution spectral observations; (ii) the instrument should be capable of making use of  the best seeing of the site as small as 0.6 $arcsecond$; (iii) the imaging Field of View (FOV) should meet the unvignetting FOV of the telescope that is 10 $\times$ 10 $arcminute$; (iv) the operating spectral range of the instrument should be 360 - 1000 $nm$; (v) the spectral resolution will be in the range 250 - 4000 covering the entire  spectral range i.e. 360-1000 $nm$ using a sets of grisms; (vi) the instrument should be designed to work on very faint stars and Galaxies ($\sim$22 mag ) at the lower spectral resolution; (vii) the instrument in polarimetric mode should be designed to use with minimal instrumental polarization effects ($\sim$0.03\%) and maximum sensitivity; (viii) the instrument should have maximum sensitivity $\sim$90\% (without telescope and CCD) and minimum light loss at the entrance pupil; (xi) the instrument size  must conform to those allowed provisions as in the instrument envelope of the  telescope's specifications;  (x) all refractive optical components will have wide band AR coating, and special designs should be taken care-off to make free from ghost images and central brightening effects.

\section{An optical layout of a focal reducer instrument}

The heart of a FOSC-type multi-mode instrument is a focal reducer that will re-image the focal plane to a suitable sized CCD camera by optimizing both the imaging scale and spectral resolution scale considering the seeing conditions of the site.  A schematic optical diagram of the instrument is shown in the figure 1. Major optical elements of the focal reducer instruments are collimator optics and camera optics. The collimator optics is located behind the telescope's focal plane and produces the parallel beam. In the collimated beam various optical components, e.g. filters (broad-band only) in a wheel, dispersing elements as grisms in a wheel and polarization optics in an $in-out$ mode, could be inserted. The narrow band filters are placed in the converging beam to avoid the central wavelength-shift in the parallel beam. The parallel beam is absolutely necessary for the dispersing elements, otherwise it will produce optical aberrations (mainly astigmatism) in an incomplete collimated beam. The collimated beam size is an important parameter that limits the resolution in the spectroscopy mode which will be discussed in later section. The camera optics reimaged the telescope's beam onto the detectors. The camera designs are needed to optimize an imaging scale without vignetting and spectral resolution scale that will be discussed in the later section.  In the focal plane of the telescope, an entrance aperture wheel is placed, which house the multiple slits with different dimensions and masks. A calibration unit is a pre-focal unit, it's optics produces the similar telescope's beam.  It is  used for wavelength calibration and flat-fielding using the different arc-lamps and continuum lamps. An auxiliary slit-viewing unit could be used to guide an astronomical objects onto the narrow slits. In table 1, the specifications of the ADFOSC instrument are listed.

\begin{figure}[htb] 
 \begin{center} 
 \includegraphics[width=4.5in]{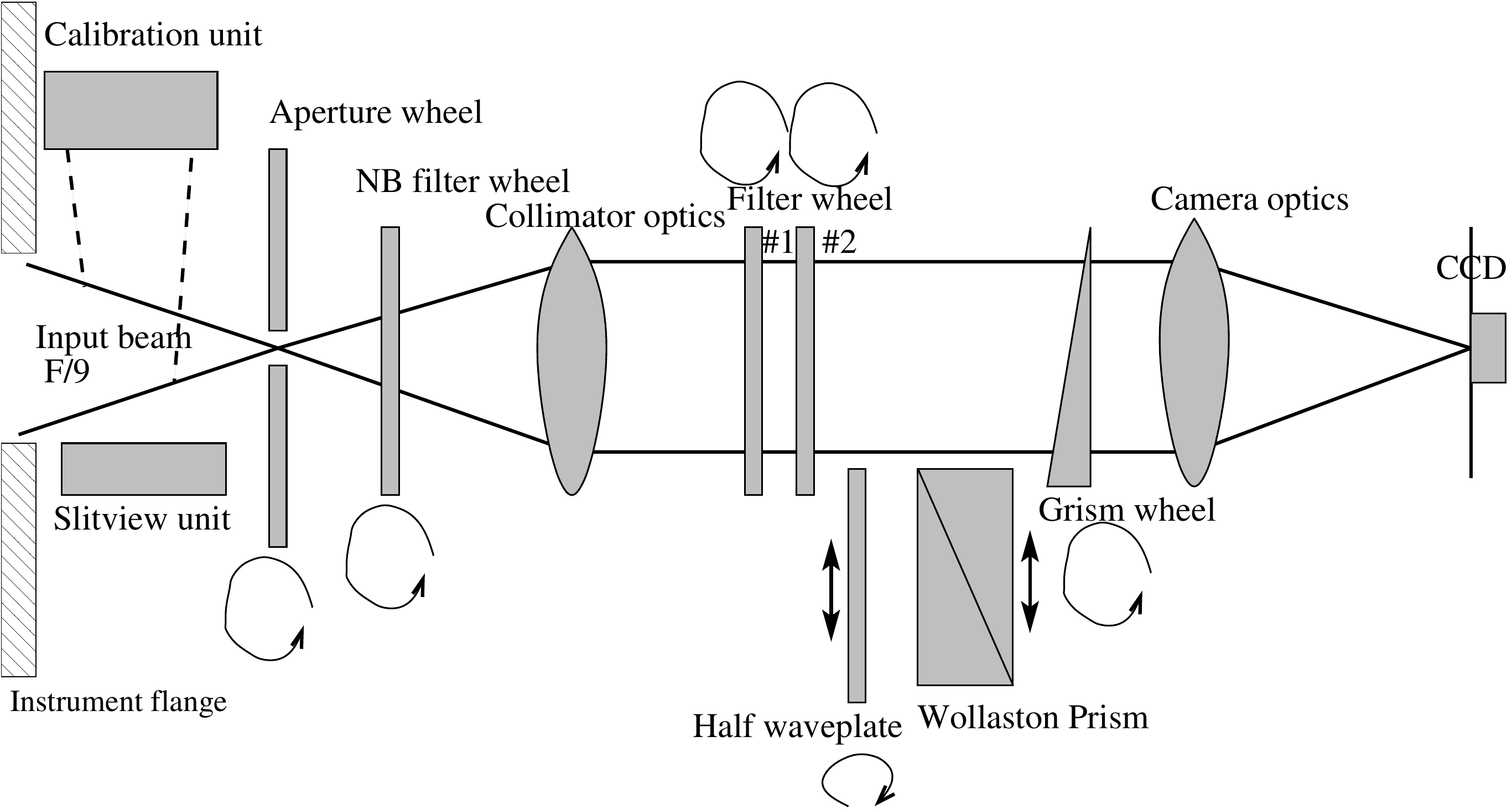} \hspace{0.1in}
 \end{center} 
 \caption{A schematic optical diagram of the ADFOSC instrument.} 
 \end{figure}

\begin{table}[thb]
\begin{center}
\caption{Specifications of ADFOSC}
\begin{tabular}{l|l}
 \hline
Mode of observations & Imaging, low-medium resolution \\
                     &spectroscopy and imaging/spctro polarimetry \\ 
Wavelength range & 360 -1000 $nm$ \\ \\
Detector array  & 2048$\times$2048 CCD with 15 $\mu$m pixel size \\ 
Imaging Field of view & 10 $\times$ 10 $arcminute$ \\ 
Imaging scale & 0.29 $arcsecond$ per 15 $\mu$m pixel \\ \\
Filters  & Broad-band : UBVRI \\  
         & Narrow-band ($nm$): 657 (H$\alpha$), 664(H$\alpha$-cont),  \\                 & 486(H$\beta$),477(H$\beta$-cont), 500 (OIII), 673 (SII) \\  \\
Spectrograph resolution &  Low-medium in the range of 250-4000 \\ \\
Entrance slits & Length ($arcmintute$) : 2, 10   \\
               & Width ($arcsecond$) : 0.5, 1, 1.2, 1.5, 2, 5, 10  \\  
               & (1 $arcsecond$ corresponds to 156 $\mu$m \\
                at the telescope's focal plane) \\ \\
Polarimeter optics & A rotating half-wave plate \\
                    & and Walloston prism \\ \\
Encircled energy & $\sim$ 80\% within 0.4 $arcsecond$ \\ 
Stray light level & less than 2\% \\ 
Instrument throughput & $\sim$90\% in imaging and polarimeter, \\
(without telescope and detector) &  in spectroscopy $\sim$60\% at peak transmission of grism \\
\hline
\end{tabular}
\end{center}
\end{table}

\section{ Imaging and Spectroscopy performance}

An efficient focal reducer spectrograph and camera requires a better pixel matching.  In a spectrograph, according to the Nyquist criterion the optimum match is two pixels per projected slit-width. In the imaging observations, imaging scale should be 2-3 pixels under best seeing conditions of the site. The median seeing of Devasthal site is  1.1 $arcsecond$ \cite{sagar,stalin}. There should be a balance between imaging scale  and projected slit-width at the detector plane to the pixel size. If the projected scale is undersampled with less than two  pixels there will be a loss in resolution for spectroscopy and a poor point spread function (psf) in imaging. While if the scale is oversampled, the per pixel signal will be less for a given exposure time, and the CCD readout noise is more significant. Thus we require a camera design to optimize the sampling as a function of seeing condition, $\phi$, of the site and the aperture D  of the telescope, the camera f-ratio for pixel matching is defined as \cite{daniel}, 

\begin{equation}
f_{camera}= \frac{4p}{D \tan \phi}
\end{equation}

where p is the pixel size of the detector, we consider here 4 pixels matching that is more favourable in many available FOSC instruments. And we assume the pixel size of 15 $\mu$m. The focal ratio of camera f$_{camera}$ calculated is shown in the table 2 for the various apertures (D) and the different seeing conditions ($\phi$).

\begin{table}[htb]
\begin{center}
\caption{Camera f-ratio for different aperture telescopes}
\begin{tabular}{cccc}
 \hline \\
\multicolumn{1}{c}{Seeing ($\phi$)}~~~~~~~ & \multicolumn{3}{c}{Aperture (D)} \\
\multicolumn{1}{c}{($arcsecond$)}      &~~~~~2.0$m$~~~~ &~~~~3.6$m$~~~~~&~~~~~8.0$m$ \\
\hline 
1.2 ~~& 5.0  & 3.0 & 1.3  \\
1.0 ~~& 6.2  & 3.4 & 1.6 \\
0.7 ~~& 9.0  & 4.9 & 2.1  \\
0.5 ~~& 12.4 & 6.8 & 3.0  \\
\hline
\end{tabular}
\end{center}
\end{table}

In table 2, it is apparent that an increase in the telescope diameter leads to more faster camera optics for a given seeing of the site. The optimal seeing value of the Astronomical site is essential for the design of the instrument optics. Thus F/3 camera optics might be suitable for our ADFOSC instrument.

\subsection{Imaging} 

The F/9 beam of the 3.6$m$ DOT  has a plate scale of 6.4 $arcsecond/mm$ at the Cassegrain focal plane. The telescope  will provide 10 X 10 $arcminute$ unvignetting imaging Field of View (FOV). The median seeing of the Devasthal site is 1.1 $arcsecond$, and the occasional best seeing is  $\sim$0.7 $arcsecond$ \cite{sagar,stalin}. The F/3 camera optics will provide the plate scale at the detector plane of 6.4$\times$9/3 = 19.2 $arcsecond/mm$, which corresponds to 0.29 $arcsecond$ per 15 $\mu$m pixel. A 2048 $\times$ 2048 pixels CCD camera will provide a FOV of 9.9 $\times$ 9.9 $arcminute$, which is fairly close to the unvignetting FOV of the telescope of 10 $\times$ 10 $arcminute$.

\subsection{Spectroscopy}

Faint object spectrographs are usually used for low-medium resolution spectroscopy using the grisms. The grisms are a combination of transmission gratings and prisms. The spectral resolution R for a particular slit-width $\phi$ is defined as \cite{daniel},

\begin{equation}
 R = \frac{d}{D \tan \phi}
\end{equation}

where d is the diameter of the collimated beam or size of dispersing elements, and D is the aperture of the telescope. To keep the resolution fixed for a given seeing value, the collimated beam diameter has to be  increased with the increasing size of the telescope. Or decreasing slit-width has the similar effect. But the optimal seeing conditions of the site is the deciding factor of the slit-width, otherwise there will be loss of signal. The 3.6m F/9 input beam size is $\sim$130 mm for a FOV of 10$\times$10 $arcmin$.  If we consider the collimated beam size of 100 $mm$, we get the  maximum resolution R of $\approx$ 4500 for a slit-width of 1.2 $arcsec$. 

The limit of resolution, $\Delta\lambda$ is defined as \cite{daniel},

\begin{equation}
\Delta\lambda = (\phi/A)(D/d)
\end{equation}

where A is the angular dispersion of a dispersing element. For a grating in Littrow configuration we have, $A = m/\sigma \cos \delta$. Where the $\sigma$ is the spacing between two successive, equally spaced grooves, and the $\delta$ is the blaze angle of the grism. The grating is used in the first-order for a maximum efficiencies, m=1.   For a few gratings from the catalog of the Richardson-Newport\footnote{http://gratings.newport.com/}, we have estimated the resolution for the ADFOSC instrument (in the table 3) with a slit-width of 1.2 $arcsec$ and the collimated beam size of 100 $mm$. 

\begin{table}
\begin{center}
 \caption{Resolution obtained using different standard gratings}
\begin{tabular}{c|c|c|c|c|c}
\hline
Grating   & Grating blaze & Blaze Angle & Resolution for & Dispersion & Catalog \\
$grooves/mm$ & wavelength ($nm$) & ($degree$) & a 1.2 $arcsec$ slit &  ($Ang./pix$) &  \\
\hline
100 & 465 & 4.6 & 225 & 5.2 & Newport \\
300 & 490 & 14.2 & 725 & 1.7 & Newport \\
600 & 540 & 34 & 1765 &  0.7 & Newport \\
\hline
\end{tabular}
\end{center}
\end{table}

\section{Summary}

We  present here the preliminary  design of a focal reducer instrument ADFOSC. The instrument will be  mounted at the Cassegrain focus of our new  3.6m telescope to be installed  at Devasthal site (near Nainital), India. The optical designs of the instrument are optimized considering the telescope parmeters and site conditions. The instrument will provide scope for  broad- and narrow-band imaging observations, spectroscopic and polarimetric observations in the wavelength region 360 -1000 $nm$.  A 2048 $\times$ 2048 CCD detector and F/3 camera optics will provide  an unvignetting FOV of the telescope (10 $\times$ 10 $arcminute$) and an imaging scale of 0.29 $arcsecond/pixel$. The maximum spectral resolution  will be, R $\approx$ 4500 for a collimator beam size of 100 $mm$ and a slit-width of 1.2 $arcsecond$. A sets of grisms having 100 -- 600 $grooves/mm$ will provide the resolutions in the range of 200 -- 2000 in the optical wavelengths.

\vspace{10pt}
\noindent Acknowledgments: The authors are thankful to Prof. S. N. Tandon (IUCAA, Pune), Prof. T.P. Prabhu (IIA, Bangalore), and  Dr. A. Chakraborty (PRL, Ahmedabad) for their valuable suggestions and discussions. The authors are extremely thankful to Prof. Ram Sagar, Director, ARIES for his constant support and encouragement for this project.

\end{document}